\documentclass[sn-mathphys-num]{sn-jnl}

\usepackage{graphicx}%
\usepackage{multirow}%
\usepackage{amsmath,amssymb,amsfonts}%
\usepackage{amsthm}%
\usepackage{mathrsfs}%
\usepackage[title]{appendix}%
\usepackage{xcolor}%
\usepackage{textcomp}%
\usepackage{manyfoot}%
\usepackage{booktabs}%
\usepackage{algorithm}%
\usepackage{algorithmicx}%
\usepackage{algpseudocode}%
\usepackage{listings}%
\usepackage{lineno}

\theoremstyle{thmstyleone}%
\theoremstyle{thmstyletwo}%

\theoremstyle{thmstylethree}%

\usepackage{ulem}

\begin{document}
\title[Artical Title]{
Understanding chiral charge-density wave by frozen chiral phonon
}

\author[1]{\sur{Shuai Zhang}}

\author*[2,3]{\sur{Kaifa Luo}}\email{kfluo@utexas.edu}

\author*[1]{\sur{Tiantian Zhang}}\email{ttzhang@itp.ac.cn}

\affil*[1]{Institute of Theoretical Physics, Chinese Academy of Sciences, Beijing 100190, China}

\affil*[2]{Department of Physics, University of Texas at Austin, Austin, Texas 78712, USA}

\affil*[3]{Oden Institute for Computational Engineering and Sciences, University of Texas at Austin, Austin, Texas 78712, USA}





\abstract{
Charge density wave~(CDW) is discovered within a wide interval in solids, however, its microscopic nature is still not transparent in most realistic materials, and the recently studied chiral ones with chiral structural distortion remain unclear.
In this paper, we try to understand the driving forces of chiral CDW transition by chiral phonons from the electron-phonon coupling scenario. 
We use the prototypal monolayer 1T-TiSe$_2$ as a case study to unveil the absence of chirality in the CDW transition and propose a general approach, i.e., symmetry-breaking stimuli, to engineer the chirality of CDW in experiments. Inelastic scattering patterns are also studied as a benchmark of chiral CDW (CCDW, which breaks the mirror/inversion symmetry in 2D/3D systems). We notice that the anisotropy changing of Bragg peak profiles, which is contributed by the soft chiral phonons, can show a remarkable signature for CCDW.
Our findings pave a path to understanding the
CCDW from the chiral phonon perspective, especially in van der Waals materials, and provides a powerful way to manipulate the chirality of CDW.}
\maketitle

\section{Introduction}

Charge density wave~(CDW) is a charge order with periodic electron density accompanied by lattice distortion, commonly identified in low-dimensional and van der Waals materials, usually followed by rearrangement of lattice structure to minimize total Coulomb energy~\cite{RevModPhys_CDW,PhysRevLett.106.106404}. 
CDW was first discovered and understood in quasi-one dimensional systems~\cite{RevModPhys_CDW, gor2012charge}, then became popular and extensively studied in layered and atomic thin transition-metal dichalcogenides (TMDs)~\cite{wilson1975charge, manzeli20172d}. For either bulk or monolayer of most TMD materials, inversion and mirror symmetries commonly exist, and are usually preserved even if CDW transitions happen.
In a few rare cases where mirrors (in 2D limit) or inversion symmetries (in 3D) are broken, chirality has to be introduced to distinguish asymmetry between left- and right-handed lattice structures. The chirality could play with a wealth of exotic phenomena ranging from charge, orbital, and magnetic order to non-trivial topological states and superconductivity. Consequently, chiral CDW~(CCDW) has unique responses to multiple external fields, such as electric fields~\cite{song2022atomic}, magnetic fields~\cite{baydin2022magnetic} and circularly polarized light~\cite{xu2020spontaneous,yang2022visualization}, and it is related to novel physical phenomena such as nonlocal Hall effect~\cite{gradhand2015optical}, optical chiral induction~\cite{xu2020spontaneous,yang2022visualization},  chiral-parity Cooper pair~\cite{hosur2013kerr,orenstein2013berry,varma2014gyrotropic,ganesh2014theoretical}, and axion insulator states~\cite{wang2013chiral,li2022chirality,you2016response}, to name a few.
It is also of great potential for electronic applications, such as the utilization of mirror domain walls in nanoelectronics and the development of CDW-based memory systems~\cite {zong2018ultrafast}.

\begin{figure}
\centering
    \includegraphics[width=0.6\textwidth]{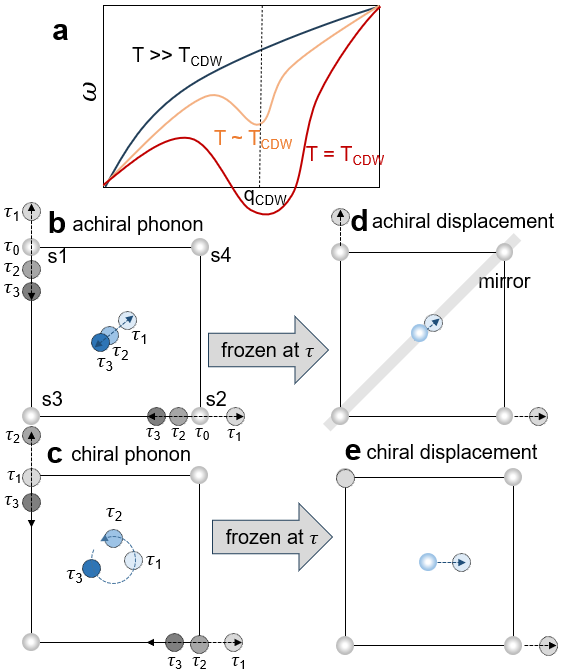}
    \caption{Mechanism for CDW and CCDW.
    (a) Phonon mode softening in the CDW process as temperature goes down. 
    (b) Atomic vibrations for an achiral phonon mode, which corresponds to linear atomic vibrations of the mass center. Grey dots are for the atoms in the unit cell at different times, $\tau_0$ (atoms located at the equilibrium positions), $\tau_1$, $\tau_2$, and $\tau_3$, while blue dots represent the motion of the mass center. 
    (c) Atomic vibration for a chiral phonon mode, which corresponds to the circular motion of the mass center.
    (d) An achiral CDW induced by the mirror-symmetric atomic displacement. 
    (e) A chiral CDW structure induced by a chiral displacement.}
    
    \label{cdw}
\end{figure}

Since the chirality information of a CDW phase is encoded in atomic positions, a wide variety of experimental techniques such as X-ray diffraction (XRD), electron diffraction, Raman spectroscopy, scanning tunneling microscope (STM)~\cite{ishioka2010chiral,xia2008polar,van2011chirality,iavarone2012evolution,castellan2013chiral,CHEN1984645,zenker2013chiral,orenstein2013berry,hosur2013kerr,varma2014gyrotropic,ganesh2014theoretical,gradhand2015optical,fu2015parity,hildebrand2018local} can be utilized to identify the CCDW. 
In general, CCDW is the result of the spontaneous breaking of the mirror (in 2D limit) or the inversion symmetry (in 3D). 
Based on this criterion, which is not restricted at first sight, 
it would be surprising that CCDW was first discovered in bulk TMD material 1T-TiSe$_2$ after over 50 years of investigation of TMDs, and only a handful of examples are found to date~\cite{neto2001charge,wang2019potential,zhang2020coherent, DFT_TiSe2_CDW,phonon_chiral_cdw_TiSe2, TA_CDW_Dai2023, NEWREF1, NEWREF2}. 
Given a bunch of other CDWs that are achiral, such as 1T-TiSe$_2$ in the monolayer limit, which is still no claim about the chiral CDW state of it, the unbalance of population implies that more careful inspections of the appearance of chirality in CDW transition are necessary.

In this work, we build a connection between CCDW and chiral phonon in the framework of strong coupling theory (i.e., electron-phonon coupling theory). The chiral displacements that entitle the CCDW are inherited from the circular motion of chiral phonons in the real space~\cite{zhang2015chiral,zhu2018observation,suri2021chiral,zhang2022chiral,ishito2023truly,zhang2023weyl, NEWREF3, NEWREF4}. To unveil the driving force of CCDW, state-of-the-art $ab$ $initio$ calculations are performed on monolayer 1T-TiSe$_2$~\cite{suzuki1984electron,neto2001charge,sipos2008mott,mulazzi2010absence,Dolui_2016,wang2022probing}, which is a prototypical member in the TMD family. Firstly, we rule out the Fermi surface nesting~(FSN) scenario for CCDW in TiSe$_2$ at DFT level, then uncover a remarkable electron-phonon coupling~(EPC) in the normal state, which indicates an EPC-induced mechanism involving soft modes. 
After establishing the microscopic mechanism, we propose that symmetry-breaking stimuli—such as polarized light, directional electric fields, magnetic fields, and mechanical strain—could serve as effective and accessible methods for selectively transforming an achiral CDW into a chiral one or switching between chiral domains. 
In addition to the commonly recognized static chiral distribution of Fourier components' intensity from STM~\cite{ishioka2010chiral,wang2022probing,Ren_2023,li2022rotation,xing2023optical} to characterize CCDW states, 
we suggest that the anisotropy of Bragg peaks' profile from XRD can serve as a dynamic indicator of CCDW. 
Our first-principles calculations for 1T-TiSe$_2$ demonstrate that the chirality angle in the thermal diffuse scattering pattern of the CCDW phase is notably larger than the angle induced by applied shear strain.


\begin{figure} 
\centering
    \includegraphics[width=0.6\textwidth]{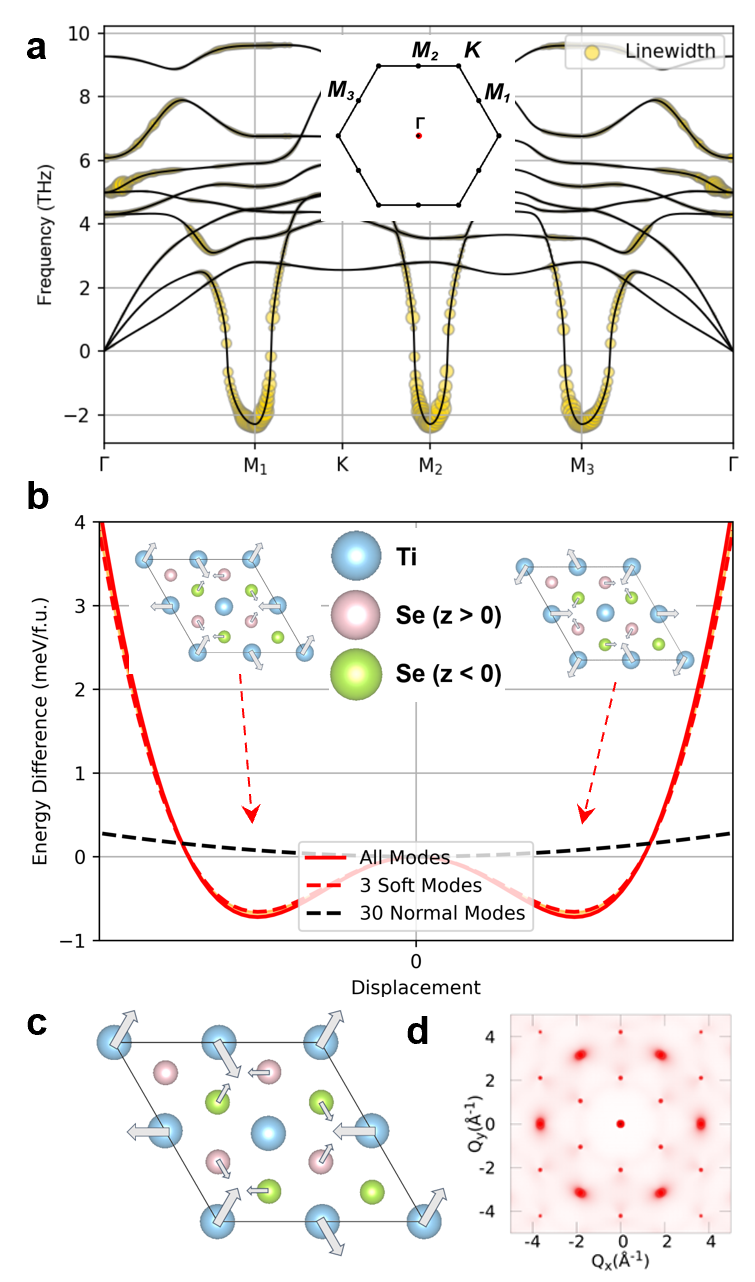}
    \caption{First-principles calculations of 2D 1T-TiSe$_{2}$.  
        (a) Phonon dispersion and linewidth (yellow dots) based on the primitive cell, showing significant electron-couplings of soft modes at three $M$. The inset shows the BZ of the primitive cell and its high-symmetry points.
        (b) Potential energy surface (PES) of 1T-TiSe$_{2}$ by linearly displacing atoms from normal state to $2\times2$ CDW state. 
        Two degenerate CDW states shown in the insets are connected by mirror symmetries. 
        The dashed double-well potential in red is formed by three frozen soft modes, while the inclusion of the other 30 normal phonon modes only slightly modifies the shape of the energy surface (solid line). (c) Atomic displacements (marked by arrows) for the $2\times2$ CDW structure. The blue, pink, and green ones represent the Ti (locating at $z=0$), Se ($z>0$), and Se ($z<0$) atoms, respectively. The main displacement of the CDW mode is contributed by the Ti atom, and the minor part is contributed by the Se atoms (the displacement vectors of Se atoms are enlarged for clearness). Each layer does not break the mirror symmetry, yet the CDW structure breaks mirror symmetry. (d) Calculated diffraction pattern of the CDW structure.}
        
    \label{fig:tise2-elph}
\end{figure}

\section{Results}

\subsection{\textbf{Connection between Chiral Phonon and Chiral CDW}}

Fermi surface nesting (FSN) and strong coupling theory are two accessible theories to understand and explain the CDW transitions in solids~\cite{han2018van}. 
FSN is a weak coupling theory generalized from the Peierls idea for the one-dimensional atomic chain, while it has been ruled out in a few typical CDW materials such as NbSe$_2$ by both the first-principles calculations and experiments, and suffering from explaining the CDW supercell size with nesting vectors in many materials~\cite{zhu2015classification, zhu2017misconceptions}.
The strong coupling theory highlights the role of EPC in CDW transitions. 
Strong EPC usually results in a strong screening of lattice dynamics, giving rise to soft phonon modes when the temperature is lower than $T_{\text{CDW}}$, as shown in Fig.~\ref{cdw} (a).
This softening, in turn, precipitates lattice instability and contributes to the CDW distorted structure.

The soft modes $\mathbf{u}(\mathbf{R},t)=\mathbf{e} \cdot e^{i(\mathbf{q}\cdot\mathbf{R}-\omega t)}$ can be further cataloged into two cases based on the atomic motion behaviors, i.e., the phonon polarization vector $\mathbf{e}$. 
We take the system shown in Fig.~\ref{cdw} (b) as an example,
in which, there are four atoms located at (1,0), (0,1), (0,0), and (1,1) in the primitive cell (which are also denoted by $s_{1...4}$.
For specifying a certain mode of atomic motion in a primitive cell, we write the phonon polarization vector on the basis of $\{u_{\kappa,\alpha}\}(\kappa=s_{1,...,4};\alpha=x,y)$.
In the case of $\mathbf{e}=(1,0,0,1,0,0,0,0)$ under the Cartesian coordinates, which corresponds to a phonon mode with two of the mirror ($M_{xy}$)-related atoms vibrate linearly and the other two atoms keep still, an achiral mode preserving mirror symmetry is obtained, as shown in Figs.~\ref{cdw} (b) and (d). This to an achiral displacement, which contributed to achiral CDW at an arbitrary time $\tau$.
In other cases where $\mathbf{e}=(1, 0, 0, e^{-i\pi/2}, 0, 0, 0, 0)$, those two non-related linearly atomic vibrations will contribute to a chiral phonon mode for the mass center, which breaks the mirror symmetry, leading to a CCDW as shown in Figs.~\ref{cdw} (c) and (e).

With the relationship between the atomic displacements of the soft modes and the CDW, one can further explore the mechanism of CCDW from the chiral phonon point of view.
To uncover the rarity of CCDW candidates in nature and offer an effective way to engineer CCDW in experiments, in the following, a careful calculation on monolayer 1T-TiSe$_2$ is used as a case study to show the mechanism for the CDW, the absence of CCDW, the way to obtain and engineer CCDW in experiments. We will utilize TiSe$_2$ to illustrate our theory.

\subsection{CDW in 1T-TiSe$_2$: relative amplitude}

Layered 1T-TiSe$_{2}$ (space group P$\bar{3}$m1) is the first system reported to host CCDW~\cite{ishioka2010chiral,xu2020spontaneous}, and it undergoes a lattice rearrangement to a $2\times 2\times 2$ supercell near $200$ K. 
One of the possible explanations of the chirality is that the locking of relative $\frac{\pi}{2}$ phases between degenerate CDWs along three orthogonal directions, and the chiral feature survives even when the system is reduced to three layers.
For monolayer 1T-TiSe$_2$, the $2\times2$ CDW phase has been proposed with $T_{\text{CDW}}=232 K$~\cite{Chen2015}, yet debate remains.
This is not immediately evident by claiming that chirality is only well-defined in 3D because CCDW in other monolayer materials has been observed, such as 1T-NbSe$_{2}$ has a CCDW structure identified by STM~\cite{SongCCDW2022}.
This anomaly inspires us to revisit CDW transition in monolayer 1T-TiSe$_{2}$.

\begin{figure}[H]
\centering
    \includegraphics[width=0.6\textwidth]{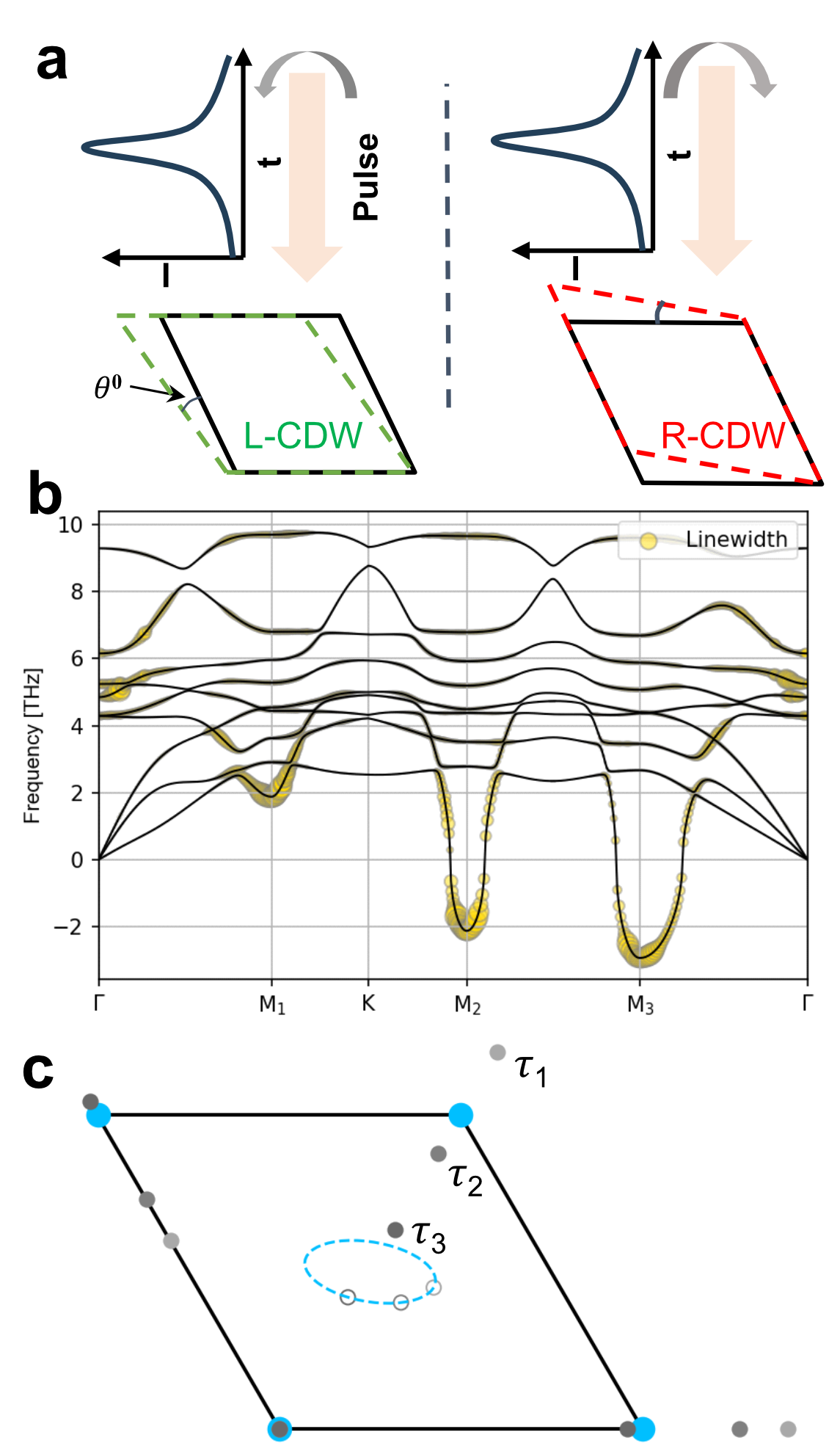}
    \caption{
    Mirror symmetry breaking external stimuli induced CCDW.
    (a) The left-handed CDW (L-CDW, green one) induced by mirror symmetry-breaking stimuli, such as a time-dependent electronic/magnetic field. The right-handed CDW (R-CDW) is likewise.
    (b) Phonon dispersion and linewidth of 1T-TiSe$_{2}$  based on the left-handed strained primitive cell. 
    The degeneracy of three soft modes at $M$ is lifted due to the breaking of the mirror and $C_3$ rotation symmetries, while their coupling strengths with electrons are still the strongest among others. 
    (c) CCDW induced by the chiral soft modes in TiSe$_2$. 
    The blue dots denote the equilibrium positions of Ti atoms for the $2\times2$ supercell (Here we admit all the Se atoms for clearness). The light-gray, gray, and dark-gray solid circles represent the positions at $\tau_1$, $\tau_2$ and $\tau_3$. The hollow circles represent the mass center of Ti atoms at these $\tau$, resulting in a chiral elliptical motion for the mass center (dashed blue line).
    } 
    \label{fig:tise2-strain2}
\end{figure}

Figure.~\ref{fig:tise2-elph} (a) shows the phonon spectra of monolayer 1T-TiSe$_{2}$, where the soft modes show the instability of the structure.
To study the CDW mechanism of the system, the Fermi nesting function $\chi_{0}''({\bf q})$ is calculated and shown in the supplementary. Excluding the artificial singularity at $\Gamma$ coming from the double-delta approximation, strong peaks of $\chi_{0}''({\bf q})$ are widely distributed around $M$, indicating that the FSN cannot quantitatively explain the selective softening of phonon modes in the vicinity of $M$, as has also been established in a few other TMD materials~\cite{zhu2015classification, zhu2017misconceptions}.

Hereafter, we turn to study the CDW mechanism under the strong coupling scenario. We calculate phonon linewidth to estimate the EPC of each mode at momentum ${\bf q}$, i.e.
\begin{equation}
    \Pi_{{\bf q}\nu}''(\omega,T)=2\pi\sum_{mn{\bf k}}|g_{mn\nu}({\bf k},{\bf q})|^{2}[f_{T}(\varepsilon_{n{\bf k}})-f_{T}(\varepsilon_{m{\bf k}+{\bf q}})]\delta(\varepsilon_{m{\bf k}+{\bf q}}-\varepsilon_{n{\bf k}}-\omega),
\end{equation}
as plotted by yellow dots in Fig.\ref{fig:tise2-elph} (a), where 
$g_{mn\nu}({\bf k},{\bf q})$ is the EPC vertex and $f_{T}(\varepsilon)$ 
is the Fermi-Dirac distribution function. 
Dominant EPCs are contributed from three in-plane acoustic phonon modes at $M$, indicating that the CDW transition may arise from anomalous strong couplings of electrons to atomic displacements in line with these phonon modes. 
Next, we move to study the CDW structure in terms of total energy and atomic displacements under the EPC scenario. In principle, identifying the structure distortion from the normal state to $2\times 2$ CDW state needs a specification of 36 degrees of freedom (12 atoms $\times$ 3 directions). 
We decompose the atomic displacements $\Delta\tau_{\kappa\alpha p}$ to $36$ phonon normal coordinates $z_{{\bf q}\nu}$ involved in the lattice reconstruction from $1\times 1$ to $2\times 2$, which shows three soft modes are the key degrees of freedom that drive CDW transition (see supplementary for details).
All the other $30$ phonon modes only slightly modify the shape of the double well and form a parabolic potential energy surface, as shown in Fig.~\ref{fig:tise2-elph} (b). 
The amplitude of the soft modes and the corresponding PES are qualitatively consistent across the different exchange-dependent (XC) functionals, as shown in the supplementary.

Three coupled order parameters that dominate the CDW state are 
\begin{equation}
    \Delta\tau_{\kappa\alpha p}^{(i)}=\sqrt{4M_{0}/N_{p}M_{\kappa}}\cos({\bf M}_{i}\cdot {\bf R}_{p}+\phi_{i}){\bf e}_{\kappa\alpha}({\bf M}_{i})|z_{i}|, 
\end{equation}
where $|z_{i}|e^{i\phi_{i}}$ is the complex normal coordinate, ${\bf R}_{p}$ is the lattice vector of unit cell $p$ in the Born von-Karman supercell, $M_{\kappa}$ is the mass of $\kappa$-th atom in a unit cell, $M_{0}$ is the proton mass as a mass reference to follow the convention, and ${\bf e}_{\kappa\alpha}({\bf M}_{i})$ are polarization vectors of soft mode at ${\bf M}_{i}$.
A linear combination of those three coupled order parameters forms the actual pattern of a CDW state, i.e., $\Delta\tau_{\kappa\alpha p}=\sum_{i=1}^{3}\Delta \tau_{\kappa\alpha p}^{(i)}$, where the amplitude $|z_{i}|$ and phase $\phi_{i}$ specifying the contribution of each order.
In 3D systems, relative phases ($\phi_{j}-\phi_{i}$) between
degenerate transverse phonon modes could share a common propagation direction, thus they can define the chirality. However, due to the absence of a common propagation direction in 2D systems, the relative phases simply relate to a global shift of a fixed point and are irrelevant to the chirality.
Thereafter, we set $\phi_{i}=0$ for simplicity, and the chirality defined by the lack of mirror symmetry we are interested in must come from the amplitude $|z_{i}|$, and $|z_1|=|z_2|=|z_3|$ restricted by $C_3$ and mirror symmetry.
Crystal structure induced by this mode preserves $C_3$ rotation symmetry while breaking the mirror symmetry, as shown in Fig.~\ref{fig:tise2-elph} (c),  yet there is no decisive experimental evidence about the chirality of the CDW phase of the monolayer TiSe$_2$, such as XRD, STM and other experimental means that can directly observe the crystal structure.

Monolayer TiSe$_2$ is composed of 3 atomic layers, Se-Ti-Se from bottom to top, and each layer has mirror symmetry, thus it is difficult for STM (which can only detect the configuration of the surface atoms) to capture mirror symmetry breaking. 
Furthermore, we performed $ab$ $initio$ calculations of diffraction pattern corresponding to XRD/electric diffraction observations~\cite{Allphonon_Marios_2021,Multiphonon_Marios_2021}, as shown by the mirror-symmetric scattering pattern in Fig.~\ref{fig:tise2-elph} (d) under 230 K. The results for the other different temperatures are shown in supplementary materials. Therefore, no mirror symmetry breaking is observed for both elastic and inelastic scattering patterns, which is also consistent with the recent experiment result~\cite{Cheng2022}.


\subsection{Engineering CCDW by shear strain}
Since neither of the two main methods can detect the chirality of the CDW phase for TiSe$_2$, a way to obtain and modulate the CCDW phase is needed.
In the recent studies concerning extant CCDW phases, there are at least four distinct external stimuli that have the potential to switch the chirality of CDW, ascertained through experimental observations~\cite{baydin2022magnetic,xu2020spontaneous,yang2022visualization,xing2023optical,nie2023unraveling,ren2023chiral,straquadine2022evidence}, including magnetic field, circularly polarized light, electric field with polarization, and local electric field pulse delivered through STM tips. 
However, altering the chirality of CDW is mostly done randomly, such as the electric pulse modulation in NbSe$_2$~\cite{SongCCDW2022}.
Here, we provide a possible method to induce the shear strain and hence the CCDW by the external electric/magnetic field, through the nonlinear electro- or magnetostrictive effects~\cite{Photostrictive_Effect_CC_2021,landau1946electrodynamics, Xing2024-ff} ( see supplementary for details), as shown in Fig.~\ref{fig:tise2-strain2} (a), which is a trigger to break the degeneracy of left-handed CCDW (L-CDW) and right-handed CCDW (R-CDW) phases.
Furthermore, the chirality of CCDW can be modulated by changing the direction of the magnetic/electric field.

Due to the breaking of vertical mirror symmetry under the perturbation, chiral phonons are expected to be obtained.
Though $M$ is the time-reversal-invariant momenta with linearly vibrating modes, the contribution for each $M$ mode will be no longer equal due to the breaking of mirror symmetry and $C_3$ rotation symmetry, as shown in Fig.~\ref{fig:tise2-strain2} (b), resulting in a soften chiral phonon mode with circular motion for the mass center, as shown in Fig.~\ref{fig:tise2-strain2} (c). The chiral phonon of the mass center contributes to the CCDW, as the case in monolayer 1T-TiSe$_2$. That is, by breaking the vertical mirror symmetry, the chiral soft mode induces a CCDW at $\tau$.
By changing the direction of the shear strain, the chirality of CCDW can be selectively switched.
In contrast to alternative methods for inducing CCDW, shear strain provides a clear advantage by a continuous modulation on the strength of the CCDW with a customizable strain strength, represented as shear strain angle $\theta^{0}$ in our case.


\begin{figure}[h]
\centering
    \includegraphics[width=0.8\textwidth]{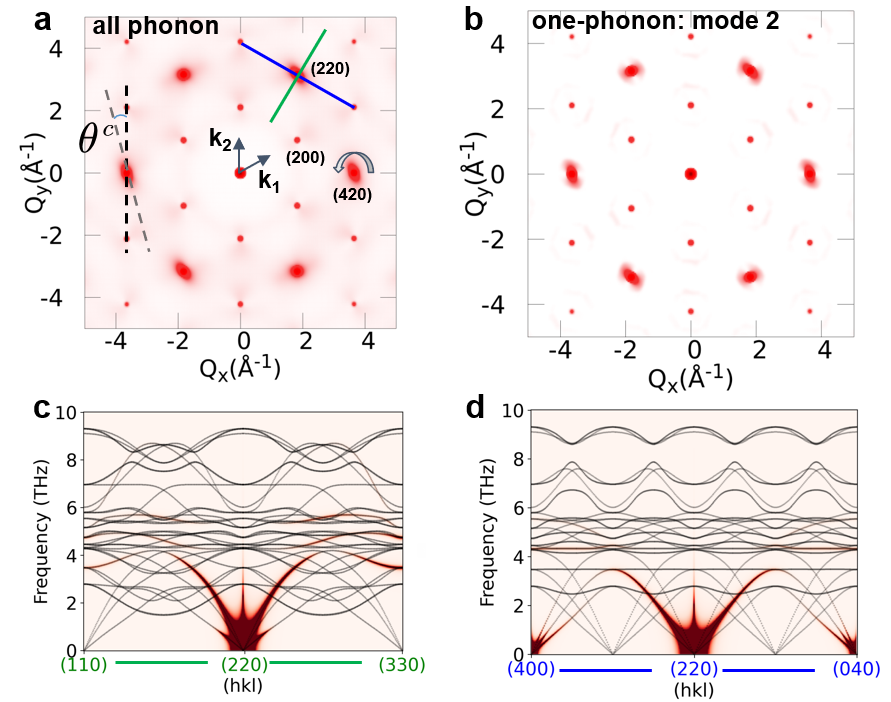}
    \caption{
    The diffuse scattering factors are calculated based on the left-handed CCDW structure with $\theta=0.05^{\circ}$.  (a) The scattering pattern with all phonon modes. The scattering pattern located at (420) is left rotated (represented by the rotation angle $\theta^c=15^\circ $), compared to the pattern in the no-sheared structure, which indicates the mirror symmetry breaking and left-chirality of the crystal structure. (b) The scattering pattern with $2nd$ phonon mode shows that the left rotated pattern at (420) is mainly contributed by the $2nd$ mode.
    (c) and (d) is the dynamic structure factors with the same structure along two paths marked in (a), passing the rotated pattern at (220). (c) DSF from (110) to (330), donated by the green line. (d) The DSF from (400) to (040), donated by the blue line. The black lines in (c) and (d) represent the phonon spectra from DFT.
    }
    \label{fig:XRD}
\end{figure}

\subsection{experimental realization and detection}
XRD/Electron diffraction are powerful method to identify the crystal structure, and hence the chirality of CDW. However, only the location or the intensity of the spots is used for identifying the crystal structures in previous studies.
We propose that the anisotropy of the Bragg peak profile can provide us with more information about the CCDW structure. 
To get a more intuitive picture, we calculated the diffraction pattern of 1T-TiSe$_2$ under 230 K, with left-hand chiral stimuli $\theta^{0}=$ 0.05$^{\circ}$, as shown in Fig.~\ref{fig:XRD} (a). Moreover, we have also calculated the diffuse scattering pattern under lower temperatures, which shows no qualitative difference (see supplementary for details).
Though the shape of the scattering pattern at $(220)$ (or $(420)$) is ellipse for both of the structures of CDW (Fig.~\ref{fig:tise2-elph} (d)) and CCDW (Fig.~\ref{fig:XRD} (a)), the latter one breaks the mirror symmetry and shows a chiral pattern.

Here we define a quantity named ``chirality angle'' ($\theta^c$) to identify the angle between the major axis of elliptical scattering spot in the CDW and CCDW phase, i.e., the degree of destruction of the mirror symmetry. The chirality angle $\theta^{c} \approx 15 ^\circ$, which is much larger than the shear angle $\theta^{0}$.
A further calculation shows that the 2$nd$ mode contributes $\theta^{c}$ most, resulting in a mirror-breaking structure, as shown in Fig.~\ref{fig:XRD} (b). Other patterns for the acoustic modes are in the supplementary.
Based on the L-CDW structure with $\theta^{c}=0.05^{\circ}$, we also calculated the dynamic structure factor (DSF, which could be seen as the diffraction with only the one phonon process), passing through $(220)$ in two paths (green and blue lines). The first one is from $(110)$ to $(330)$, as shown in in Fig.~\ref{fig:XRD} (c), while the second one is from $(400)$ to $(040)$, as shown in Fig.~\ref{fig:XRD} (d).
Both (b) and the blue cut in (d) show that the 2$nd$ mode contributes the most, thus the change of intensity along the major axis is mainly contributed by the 2$nd$ phonon mode. 
The chirality angle is found to be qualitatively robust by using different XC functionals in the first-principles simulations. Details for these features and their dependencies are shown in supplementary materials.

\section{Discussion}
In this work, we pave a path to understanding the CCDW from the chiral phonon perspective, with the primary aim of establishing the connection between CCDW, chiral displacement, and chiral phonons under the EPC scenario.
We employ 2D 1T-TiSe$_2$ as a case to elucidate the mechanisms governing the CDW and the rationale behind the absence of CCDW in this material.
Furthermore, we also propose that shear strain or equivalent methods can be a way to obtain and modulate CCDW, which can be realized in multiple experimental ways, such as by electric/magnetic fields, circularly polarized light, mechanical strain, and so on. 
Our scenario of electron-chiral phonon coupling induced CCDW can be generalized to 3D systems, such as perovskites, and applicable to both metals and insulators.
We also propose that the anisotropy of the inelastic scattering intensity can be a benchmark for CCDW, which can be verified by X-ray/electron beam.

\section{Methods}

\subsection{Ground-state calculations}
\textit{Ab initio} calculations were carried out for 1T-TiSe$_{2}$ (space group $P\bar{3}m1$), using the relaxed in-plane lattice parameter $a=3.442$ $\mathring{\text{A}}$, and interlayer distance $c=3.762a$. The Titanium atom is located at $1a$ position (0,0,0), and two Senium atoms are located at $2d$ $(1/3, 2/3, 0.443a)$. We used DFT within the generalized gradient approximation of PW91 LDA functionals~\cite{PW91}. The core–valence interaction was described using norm-conserving pseudopotentials. Electron wavefunctions were expanded in a plane waves basis set with a kinetic energy cutoff of 100 Ry, and the Brillouin zone was sampled using a 24×24×1 $\Gamma$-centered Monkhorst–Pack mesh. Lattice-dynamical properties were calculated using density functional perturbation theory (DFPT)~\cite{dfpt_rev}. All DFT and density functional perturbation theory calculations were performed using the Quantum ESPRESSO package~\cite{Giannozzi_2009, Giannozzi_2017}. We adopt a degauss $\sigma=$1.9 mRy of Fermi-Dirac smearing function in total energy calculations. Note that, the soft modes can be stabilized by increasing degauss to 4 mRy.

\subsection{Electron-phonon coupling}
Calculations of electron-phonon couplings were performed using the EPW code~\cite{Ponce2016}. To evaluate phonon self-energy and linewidth, we computed electronic and vibrational states as well as the scattering matrix elements on a 24×24×1 and 4×4×1 Brillouin-zone grid, respectively. These quantities were interpolated with \textit{ab initio} accuracy onto a fine grid with 200×200×1 $k$-points and a high-symmetry path of $q$-points using EPW. Temperature effects were accounted for by including the Fermi–Dirac occupation in the spectral function, corresponding to 50K below the CDW transition temperature.

\subsection{Phonon Diffused scattering calculation}
The multi-phonon diffused scattering method~\cite{Allphonon_Marios_2021,Multiphonon_Marios_2021}, which is implemented in the EPW package suite~\cite{Ponce2016}.
For both the CDW and CCDW states, we have chosen a $40\times 40 \times 1$ supercell for getting a converged scattering pattern.
For the $2\times2$ CDW and CCDW structures, the dynamic matrix is calculated on a $2 \times 2 \times 1$ $q$ mesh, with the electronic self-consistent field calculation implemented on a $12 \times 12 \times 1$ mesh.

\section{Data Availability}
The original data used in this work are available from the corresponding authors upon reasonable request.

\section{Code Availability}
The code used in this work is available from the corresponding authors upon reasonable request.

\section{Acknowledgement}
We acknowledge the support from the National Natural Science Foundation of China (Grant Nos. 12047503 and 12374165), and National Key R\&D Young Scientist Project 2023YFA1407400.

\section{Author Contributions}
T.Z. and K.L. devised the project idea and prepared the main part of the manuscript. S.Z. and K.L. performed the first-principles calculations. All the authors analyzed and discussed the results and edited the manuscript.

\section{Competing Interests}
The authors declare no competing financial or non-financial interests.

\bibliography{main}

\section{Figure Legends}
\subsection{Legend of Fig.~\ref{cdw}}
Mechanism for CDW and CCDW.
    (a) Phonon mode softening in the CDW process as temperature goes down. 
    (b) Atomic vibrations for an achiral phonon mode, which corresponds to linear atomic vibrations of the mass center. Grey dots are for the atoms in the unit cell at different times, $\tau_0$ (atoms located at the equilibrium positions), $\tau_1$, $\tau_2$, and $\tau_3$, while blue dots represent the motion of the mass center. 
    (c) Atomic vibration for a chiral phonon mode, which corresponds to the circular motion of the mass center.
    (d) An achiral CDW induced by the mirror-symmetric atomic displacement. 
    (e) A chiral CDW structure induced by a chiral displacement.

  \subsection{Legend of Fig.~\ref{fig:tise2-elph}}
  First-principles calculations of 2D 1T-TiSe$_{2}$.  
        (a) Phonon dispersion and linewidth (yellow dots) based on the primitive cell, showing significant electron-couplings of soft modes at three $M$. The inset shows the BZ of the primitive cell and its high-symmetry points.
        (b) Potential energy surface (PES) of 1T-TiSe$_{2}$ by linearly displacing atoms from normal state to $2\times2$ CDW state. 
        Two degenerate CDW states shown in the insets are connected by mirror symmetries. 
        The dashed double-well potential in red is formed by three frozen soft modes, while the inclusion of the other 30 normal phonon modes only slightly modifies the shape of the energy surface (solid line). (c) Atomic displacements (marked by arrows) for the $2\times2$ CDW structure. The blue, pink, and green ones represent the Ti (locating at $z=0$), Se ($z>0$), and Se ($z<0$) atoms, respectively. The main displacement of the CDW mode is contributed by the Ti atom, and the minor part is contributed by the Se atoms (the displacement vectors of Se atoms are enlarged for clearness). Each layer does not break the mirror symmetry, yet the CDW structure breaks mirror symmetry. (d) Calculated diffraction pattern of the CDW structure.

\subsection{Legend of Fig.~\ref{fig:tise2-strain2}}
    Mirror symmetry breaking external stimuli induced CCDW.
    (a) The left-handed CDW (L-CDW, green one) induced by mirror symmetry-breaking stimuli, such as a time-dependent electronic/magnetic field. The right-handed CDW (R-CDW) is likewise.
    (b) Phonon dispersion and linewidth of 1T-TiSe$_{2}$  based on the left-handed strained primitive cell. 
    The degeneracy of three soft modes at $M$ is lifted due to the breaking of the mirror and $C_3$ rotation symmetries, while their coupling strengths with electrons are still the strongest among others. 
    (c) CCDW induced by the chiral soft modes in TiSe$_2$. 
    The blue dots denote the equilibrium positions of Ti atoms for the $2\times2$ supercell (Here we admit all the Se atoms for clearness). The light-gray, gray, and dark-gray solid circles represent the positions at $\tau_1$, $\tau_2$ and $\tau_3$. The hollow circles represent the mass center of Ti atoms at these $\tau$, resulting in a chiral elliptical motion for the mass center (dashed blue line).

\subsection{Legend of Fig.~\ref{fig:XRD}}
    The diffuse scattering factors are calculated based on the left-handed CCDW structure with $\theta=0.05^{\circ}$.  (a) The scattering pattern with all phonon modes. The scattering pattern located at (420) is left rotated (represented by the rotation angle $\theta^c=15^\circ $), compared to the pattern in the no-sheared structure, which indicates the mirror symmetry breaking and left-chirality of the crystal structure. (b) The scattering pattern with $2nd$ phonon mode shows that the left rotated pattern at (420) is mainly contributed by the $2nd$ mode.
    (c) and (d) is the dynamic structure factors with the same structure along two paths marked in (a), passing the rotated pattern at (220). (c) DSF from (110) to (330), donated by the green line. (d) The DSF from (400) to (040), donated by the blue line. The black lines in (c) and (d) represent the phonon spectra from DFT.

\end{document}


\title{
Supplementary information of ``Understanding chiral charge-density wave by frozen chiral phonon''
}

\author{Shuai Zhang}
\affiliation{Institute of Theoretical Physics, Chinese Academy of Sciences, Beijing 100190, China}

\author{Kaifa Luo}
\email{kfluo@utexas.edu}
\affiliation{Department of Physics, University of Texas at Austin, Austin, Texas 78712, USA}
\affiliation{Oden Institute for Computational Engineering and Sciences, University of Texas at Austin, Austin, Texas 78712, USA}

\author{Tiantian Zhang}
\email{ttzhang@itp.ac.cn}
\affiliation{Institute of Theoretical Physics, Chinese Academy of Sciences, Beijing 100190, China}


\maketitle

\section{Fermi-surface nesting calculation}
The Fermi surface~(FS) of monolayer 1T-TiSe$_{2}$ consists of a $\Gamma$-centered hole pocket from $3d$ orbitals of titanium and three $M$-centered electron pockets from $4p$ orbitals of selenium. In the $2\times2$ supercell, electron pockets will be folded to the BZ center and overlap with the hole pocket, as shown in Supplementary Figure~\ref{fig:fsn}. All coincidence of electron-hole pairs in this reconstructed FS will screen ion-ion interaction and renormalize phonon dispersions. To estimate nesting strength, we turn to the Fermi nesting function, the imaginary part of static susceptibility functions $\chi_{0}({\bf q})$ with double-delta approximation (as shown in Supplementary References~\cite{PhysRevB.82.184509}), 
$\chi_{0}''({\bf q})=\sum_{{\bf k}}\delta(\varepsilon_{{\bf k}}-\varepsilon_{F})\delta(\varepsilon_{{\bf k}+{\bf q}}-\varepsilon_{F})$. As shown in  Supplementary Figure~\ref{fig:fsn} (b), strong peaks of $\chi_{0}''({\bf q})$ around $M$ indicate strong screening to lattice dynamics. To exam the images of these nesting on phonon dynamics, we calculate phonon dispersions $\omega_{{\bf q}\nu}$ in Supplementary Figure~2 (a) of the main text. If one overlay $\chi_{0}''({\bf q})$ and $\omega_{{\bf q}\nu}$, the swiggy shape of  $\chi_{0}''({\bf q})$ cannot explain the exclusively softening of phonon modes near $M$. Therefore, as has been established in other TMD materials, it is questionable to understand CDW transition with FSN.

\begin{figure}[h]
    \includegraphics[width=0.4\textwidth]{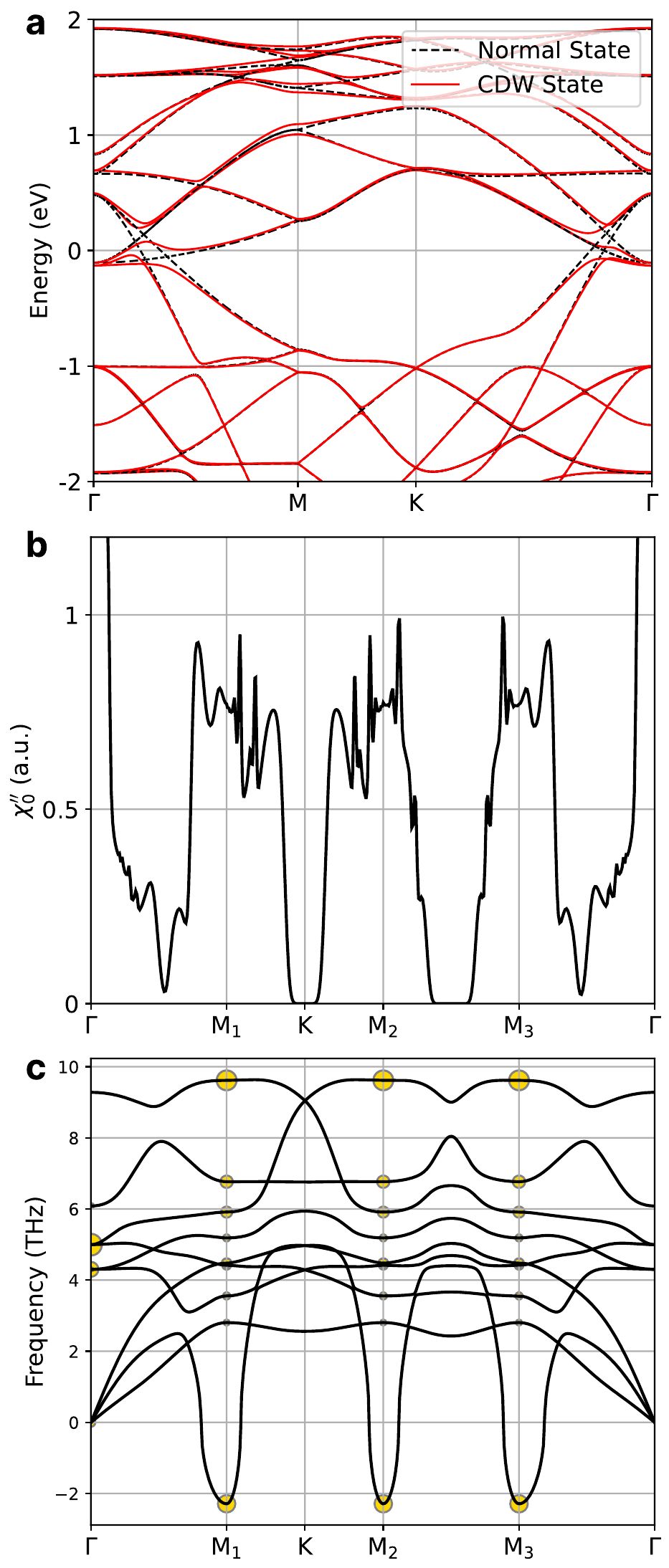}
    \caption{
    (a) Band structure of $2\times 2$ supercell 1T-TiSe$_{2}$, in CDW phase (red) and normal phase (black), respectively. 
    (b) Nesting function.
    (c) Decomposition of atomic displacements in the CDW ground state to all phonon modes, {which shows three soft modes at $M$ marked by the yellow dots dominate the formation of the CDW supercell.}
    }
    \label{fig:fsn}
\end{figure}

\section{Phonon Projection}

By projecting $\Delta\tau_{\kappa\alpha p}$ to three soft modes and linearly interpolating the atomic displacements, the double-well feature of energy difference is reproduced, and only 8.7\% shallower than the full potential energy surface.

\begin{figure}[H]
    \includegraphics[width=0.45\textwidth]{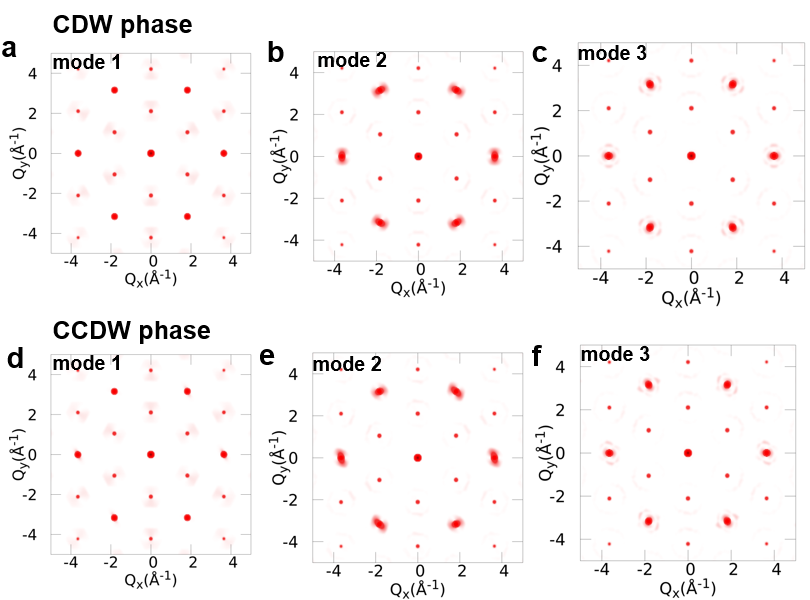}
    \caption{Phonon mode resolved scattering pattern calculated based on the (a-c) CDW and (d-f) CCDW phase, where (a) - (b) ((d) - (f)) represents the $1st$ to $3rd$ phonon mode for the CDW (CCDW) phase, respectively.}
    \label{fig:mode-xrd}
\end{figure}

\begin{figure}[H]
    \includegraphics[width=0.45\textwidth]{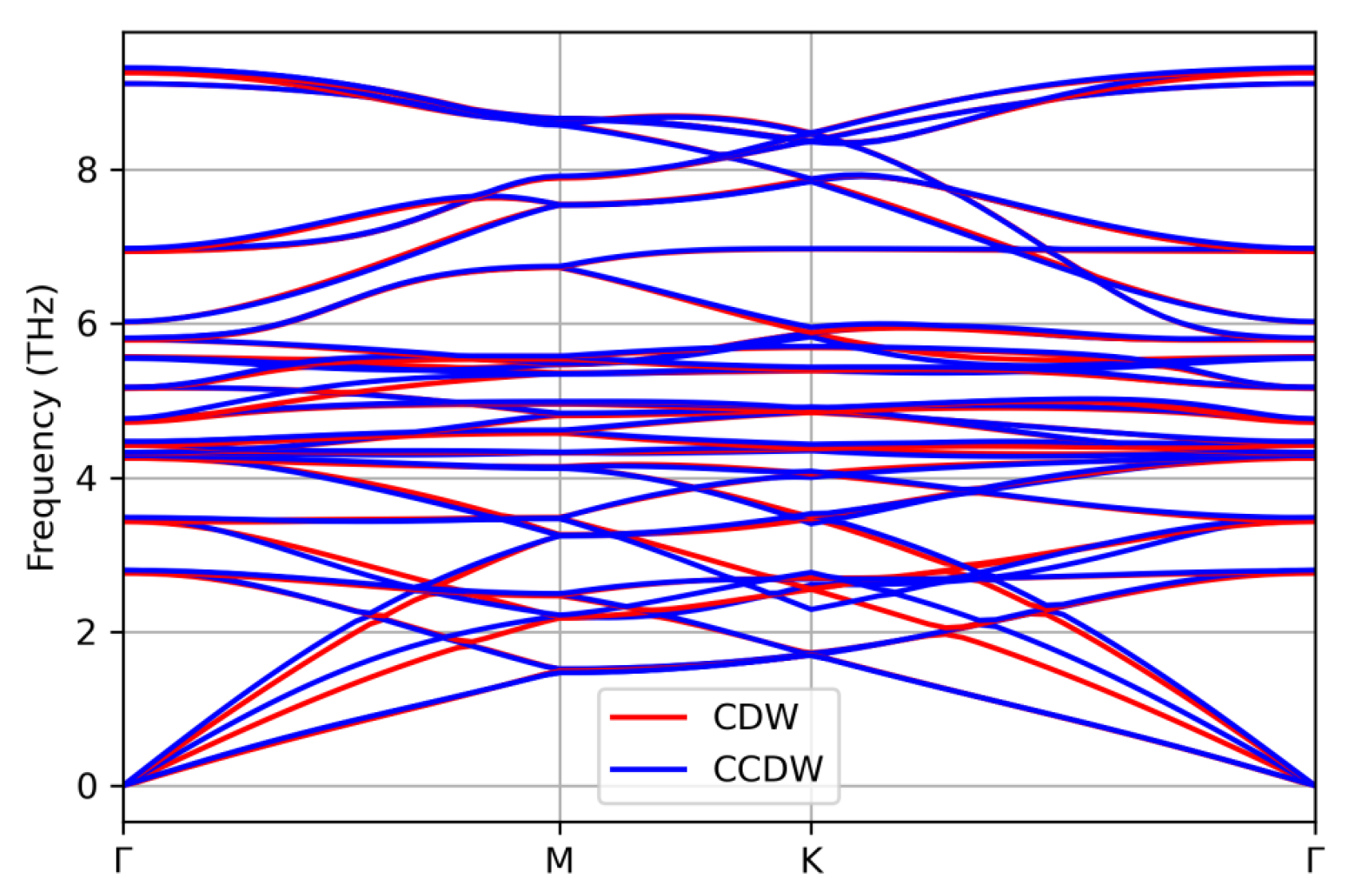}
    \caption{The phonon spectra of the CDW and CCDW states.}
    \label{fig:ph_CDW}
\end{figure}

\section{Phonon Diffused scattering calculation }
Except for the scattering pattern, we also performed a mode-resolved scattering pattern for both structures, as shown in Supplementary Figure~\ref{fig:mode-xrd}, which includes the three first phonon modes. By comparing the results of both phases, we can conclude that the local rotation of the scattering pattern is mainly contributed by the $2nd$ mode.
Furthermore, the phonon spectra of the CDW/CCDW state are shown in Supplementary Figure~\ref{fig:ph_CDW}.

\begin{figure*}[]
  \centering
  
  \includegraphics[width=0.8\textwidth]{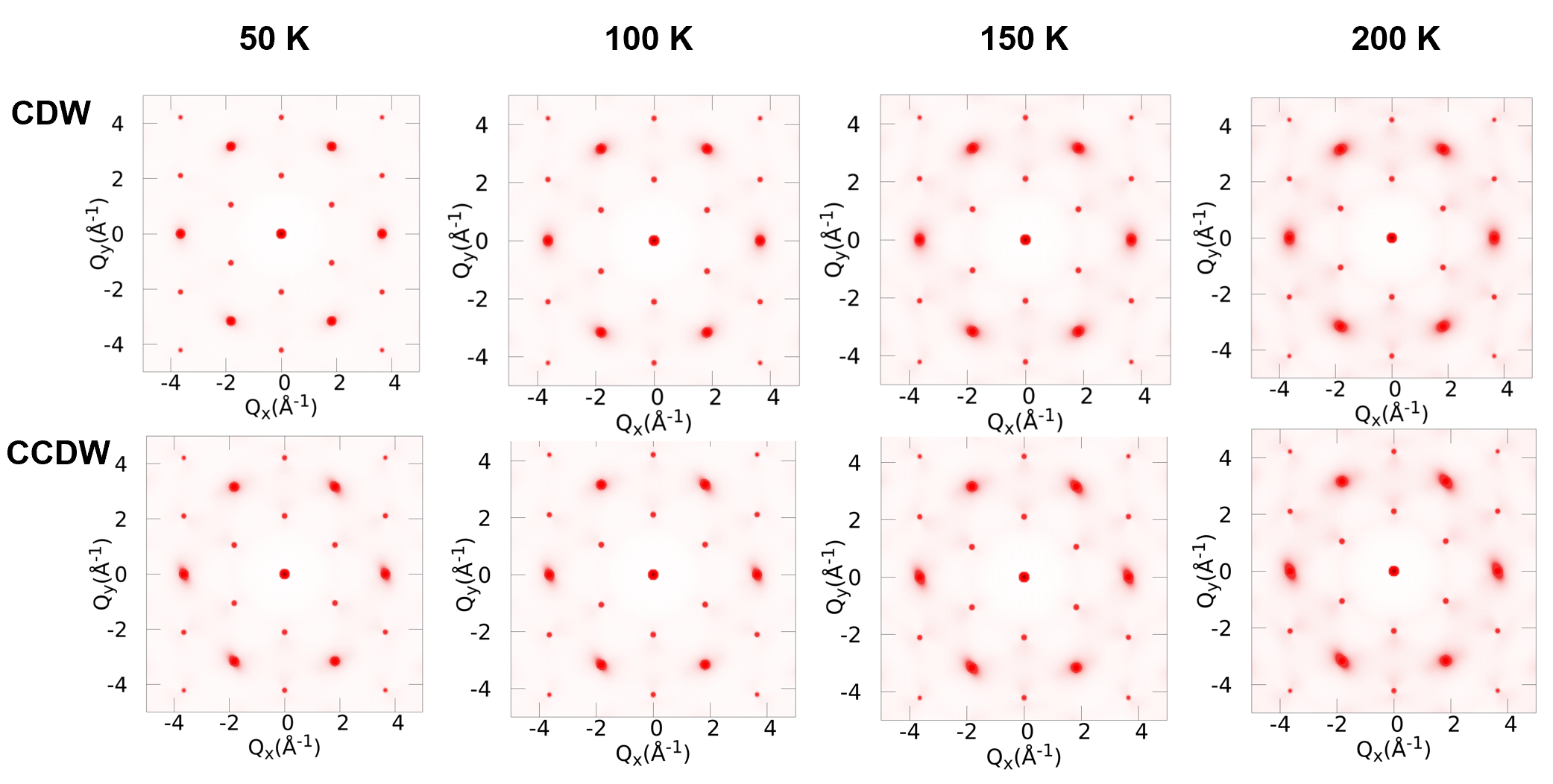}
  \caption{\label{fig:LowTXRD}
 The pattern of thermal diffusion scattering for CDW and CCDW state at different temperatures.}
\end{figure*}

To get the influence of temperature on the pattern of phonon diffuse scattering, we also calculated the result at 50 K, 100 K, 150 K, and 200 K for both of the CDW and CCDW states, as shown in Supplementary Figure~\ref{fig:LowTXRD}.
From which we can tell that there is no qualitative difference from the conclusion in the main text. As temperature goes down, only the spread of the pattern shrinks.

\section{Experimental Proposals of Shear Deformation}

For small strains, the deformed lattice vectors are linear transformations by a strain tensor:
\begin{equation}
    a_{i\alpha}'=(\delta_{\alpha\beta}+\varepsilon_{\alpha\beta})a_{i,\beta}.~(i=1,2;~\alpha,\beta=x,y)
\end{equation}
To achieve a left-handed lattice deformation presented as Supplementary Figure 3(a) in the manuscript, up to a rigid-body rotation of the lattice, the strain tensor reads
\begin{equation}
    (\varepsilon_{\alpha\beta})=\varepsilon_{xy}\begin{pmatrix}
    0 & 1\\
    1 & 2\sqrt{3}
    \end{pmatrix}
\end{equation}
with a single independent component $\varepsilon_{xy}<0$. The diagonal and off-diagonal components correspond to uniaxial and shear strains, which are routinely applied in 2D materials engineering nowadays.

Alternative approaches to induce strain are by applying external fields. A strain can be induced through piezoelectric effects. Combined with the photovoltaic effect which transforms external light into the electric field, the photostrictive effect can be expected to link light to mechanical strain. 
The photostrictive effect is relatively exotic, but it has been discovered in various systems including ferroelectrics and polar semiconductors. A recent review paper by Chen and Yi (Supplementary References ~\cite{Photostrictive_Effect_CC_2021}) discussed the mechanisms and applications of photostrictive effects from experimental aspects. 
A benefit of photo-striction is the ultra-fast chirality switching. 
Because the transition duration of CDW states from normal states is usually comparable with periodicities of phonons (1$\sim$10 picoseconds), the switching speed of chirality in CCDWs induced by oscillating light could be around terahertz, which is 2-3 orders of magnitude faster than the writing or reading speeds of commercial RAMs ($\sim$0.1GB/s) or SSDs ($\sim$0.5GB/s) for data storage.

Unfortunately, for common high-symmetric 1T- (no.164 space group) and 2H-phase (no.194 space group) of monolayer TMDs, such as 1T-TiSe$_{2}$ in this study, piezoelectric coupling are forbidden by symmetries. Accordingly, the photostriction tensor is forbidden as well. Although these two effects could be potential mechanisms to generate strain in lower-symmetry systems such as 1T'-phase TMDs, they are excluded in 1T-TiSe$_{2}$.

As the linear coupling between electric fields and strain is forbidden, their next nonlinear coupling, electro-striction, becomes the leading effect. The strain is generated by the square of electric fields $E_{i}$ through an electro-striction tensor $\gamma_{ijkl}$:
\begin{equation}
    \varepsilon_{ij}=\gamma_{ij,kl}E_{k}E_{l}
\end{equation}
in which $\gamma_{xx,xx}$, $\gamma_{xx,yy}$, $\gamma_{xy,xz}$ and $\gamma_{xy,xy}$ are symmetry-allowed. Similarly, the nonlinear coupling between magnetization $M_{i}$ and strain are also allowed and obey the same symmetry constraints, their relation reads
\begin{equation}
    \varepsilon_{ij}=N_{ij,kl}M_{k}M_{l}
\end{equation}
through the magneto-striction tensor $N_{ij,kl}$. As results, the in-plane external electric or magnetic fields and the substrate-induced electric field would serve as the sources of shear strain $\varepsilon_{xy}$ to deform the lattice. The strong electric and magnetic fields in a laser could also induce temporal strain and ultra-fast switch of chirality in CCDWs.

\begin{figure}[]
  \centering
  
  \includegraphics[width=0.5\textwidth]{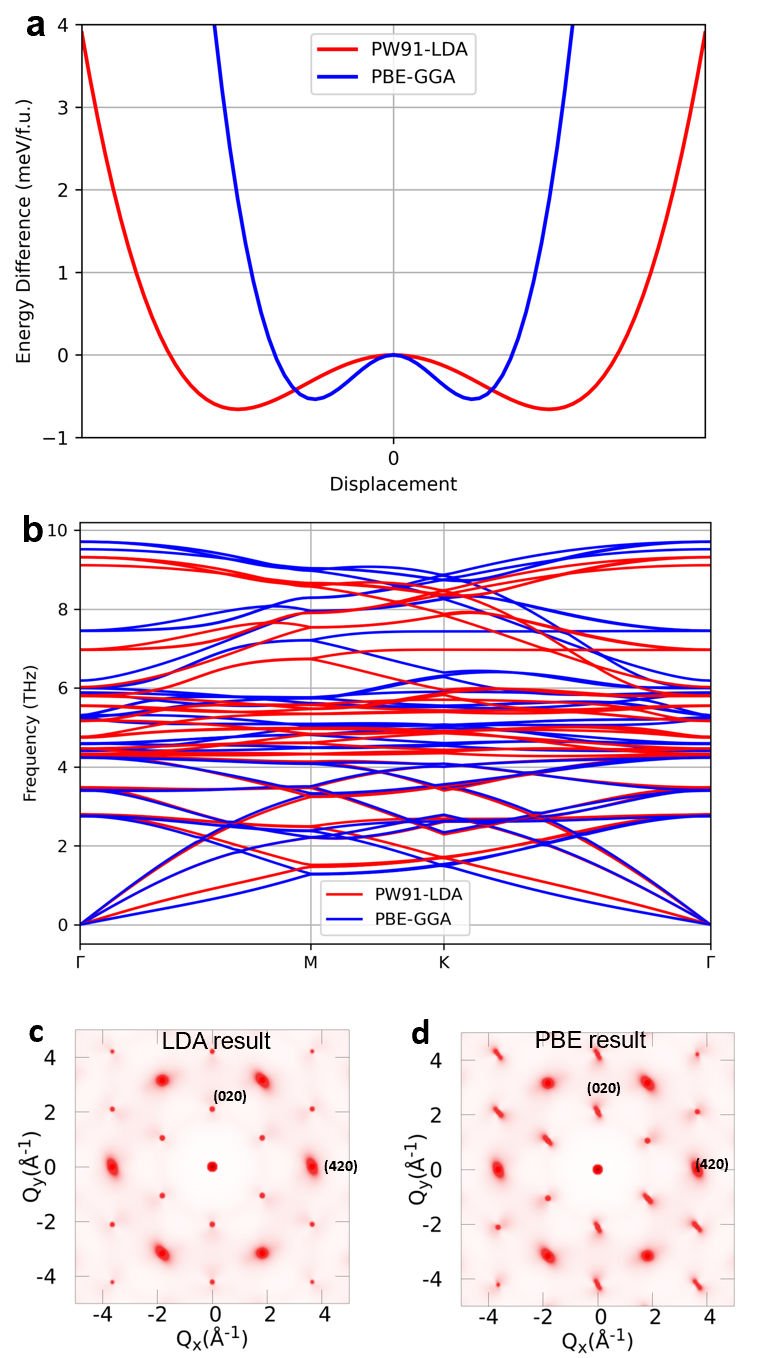}
  \caption{\label{fig:CCDW_LDAvsPBE}
(a) The potential energy surface and (b) the phonon spectra calculated with different functionals in the CCDW phase. (c-d) Thermal diffusion scattering patterns for CCDW state with the PW91-LDA and PBE-GGA functionals, respectively.}
\end{figure}

\subsection{Exchange-correlated functional dependency}

{To illustrate the sensitivity of our result to different types of exchange-correlation (XC) functionals, we calculated the potential energy surface (PES), phonon spectra, and the inelastic scattering patterns based on the PW91-LDA and PBE-GGA functionals with the $2\times2$ CCDW structure, as shown in Supplementary Figure~\ref{fig:CCDW_LDAvsPBE}. 
The fully relaxed lattice constants for LDA and PBE are 3.44 $\mathring{\text{A}}$ and 3.54 $\mathring{\text{A}}$, respectively.}

{For the PES shown in Supplementary Figure~\ref{fig:CCDW_LDAvsPBE} (a), the depth of the double well calculated with PBE functional is shallower than the one calculated with LDA functional, yet both of them have the ``double-well'' features.
}
{The phonon spectra calculated with different functionals in the CCDW phase are also different, as shown in Supplementary Figure~\ref{fig:CCDW_LDAvsPBE} (b), where the acoustic phonon with PBE functional is slightly lower than the ones with LDA functional.
Such difference can be attributed to the negative pressure-analogous effect caused by the GGA functionals, under which the lattice constant expands~\cite{Favot1999_ph_LDA_PBE}.
By comparing the inelastic scattering patterns shown in Supplementary Figure.~\ref{fig:CCDW_LDAvsPBE} (c) and (d), one may notice that both the intensity and the anisotropy for the pattern with the largest intensity located at $(420)$ are almost unchanged under different functionals. Thus, the chirality angle is a robust feature for CCDW.}

\newpage

\bibliography{supp}